# Encryption scheme based on Automorphism Group of Hermitian Function Field with Homomorphic Encryption

Gennady Khalimov[1][0000-0002-2054-9186], Yevgen Kotukh [2][0000-0003-4997-620X]

[1]Kharkiv National University of Radioelectronics, Kharkiv, 61166, Ukraine hennadii.khalimov@nure.ua
[2]Yevhenii Bereznyak Military Academy, Kyiv, Ukraine yevgenkotukh@gmail.com

**Abstract.** This article proposes a comprehensive approach to implementing encryption schemes based on the automorphism group of the Hermitian function field. We utilize a three-parameter group with logarithmic representations outside the group's center. In this work, we enhance the Hermitian function field-based encryption scheme with homomorphic encryption capabilities, which constitutes a significant advantage of our implementation. Both the attack complexity and the encrypted message size are directly correlated with the order of the group.

## INTRODUCTION

Advances in large-scale quantum computing present significant security challenges for widely-used public-key cryptosystems. Shor's quantum algorithm for solving integer factorization and discrete logarithm problems renders cryptosystems such as RSA and ECC vulnerable.

The utilization of hard-to-solve group-theoretic problems represents a promising direction for constructing quantum-resistant cryptosystems [1-4]. Wagner and Magyarik [5] initially proposed using the unsolvable word problem in permutation groups for quantum-stable cryptographic constructions. The quantum security of such approaches depends on specific implementation details, with Grover's quantum algorithm serving as a preliminary basis for cryptanalysis.

The unsolvable word problem concept was first implemented in cryptosystems utilizing logarithmic signatures [6]. These signatures, representing specialized factorizations applicable to finite groups, underwent several improvements [7-8]. The most recent iteration, known as MST3 [8], is based on the Suzuki group.

In 2008, Magliveras et al. [9] identified limitations in using periodic logarithmic signatures within the MST3 framework. Subsequently, Svaba et al.

[10] enhanced security through the eMST3 cryptosystem by incorporating a secret homomorphic cover. Further advancements came in 2018 when T. van Trung [11] proposed an MST3 approach utilizing strong aperiodic logarithmic signatures for Abelian p-groups. A comprehensive analysis by Y. Cong et al. [12] established MST3 as a viable post-quantum cryptographic candidate.

While the original MST3 construction was based on the Suzuki group, several approaches have been explored to enhance its security and efficiency [13-16]. A particularly valuable development involves improving encryption secrecy through large-order multiparameter groups while optimizing computational efficiency over small finite fields. Researchers demonstrated the applicability of this approach for calculating logarithmic signatures outside the group center, utilizing automorphisms of groups over Suzuki, Hermitian of Ree functional fields with large orders. In [13,16], a three-parameter automorphism group over the Hermitian functional field was first proposed for constructing quantum-stable encryption schemes. However, early implementations suffered from weak cohesion of logarithmic signature keys, leading to vulnerability against sequential key recovery attacks with breaking complexity equal to the finite field size.

This paper proposes a new secure encryption scheme based on the automorphism group of the Hermitian function field enhanced with homomorphic encryption techniques.

## THE AUTOMORPHISM GROUP OF THE HERMITIAN FUNCTION FIELD

The Hermitian function field $Herm | F_{q^2}$ in [13] is defined to be the function field over $F_{q^2}$ generated by two elements $x, y$ such that $y^q + y = x^{q+1}$.

We use $Aut(Herm)$ of the $Herm | F_{q^2}$ that can be presented as follows

$H := Aut(Herm) = \{\psi : Herm \mapsto Herm | \psi$ of $Herm | F_{q^2}\}$ and it's extremely large.

This group has the order equal to $ordA = q^3(q^2-1)(q^3+1)$. The decomposition group $H(P_\infty)$ has got all $Aut(Herm)$ of $Herm|F_{q^2}$ the following properties

$$\begin{cases} \psi(x) = \alpha x + \beta \\ \psi(y) = \alpha^{q+1} y + \alpha\beta^q x + \gamma, \end{cases}$$

where $\alpha \in F_{q^2}^* := F_{q^2} \setminus \{0\}$, $\beta \in F_{q^2}$ and $\gamma^q + \gamma = \beta^{q+1}$. It has order $ordH(P_\infty) = q^3(q^2-1)$.

Here we have structure for the group as follows

$$[\alpha_1, \beta_1, \gamma_1] \cdot [\alpha_2, \beta_2, \gamma_2] = [\alpha_1\alpha_2, \alpha_2\beta_1 + \beta_2, \alpha_2^{q+1}\gamma_1 + \alpha_2\beta_2^q\beta_1 + \gamma_2].$$

Thus, we have an identity $[1,0,0]$, and the inverse of $[\alpha, \beta, \gamma]$ is

$$[\alpha, \beta, \gamma]^{-1} = [\alpha^{-1}, -\alpha^{-1}\beta, \alpha^{-(q+1)}\gamma^q].$$

$H(P_\infty)$ can be represented more simply by due to the off characteristic:

$$H(P_\infty)\left\{\left[\alpha, \beta, \frac{\beta^{q+1}}{2} + \gamma\right] \middle| \alpha \in F_{q^2}^*, \beta \in F_{q^2}, \gamma^q + \gamma = 0\right\}.$$

The unique $p$-Sylow subgroup of $H(P_\infty)$ can be designated by $H_1(P_\infty)$ within the following representation $H_1(P_\infty) = \{\psi \in H(P_\infty) | \psi(x) = x + \beta$ for some $b \in F_{q^2}\}$.

In such a case we have an order equal to $q^3$ for the unique $p$-Sylow subgroup of

$$\begin{cases} \psi(x) = x + \beta \\ \psi(y) = y + \beta^q x + \gamma, \end{cases}$$

where, $\beta \in F_{q^2}$, $\gamma^q + \gamma = \beta^{q+1}$ and its order equal to $q^3$ as we mentioned above. The structure for the group can be achieved by the subgroup of PGL(3,k) presentation:

$$H_1 := \left\{\begin{pmatrix} 1 & \beta & \gamma \\ 0 & 1 & \beta^q \\ 0 & 0 & 1 \end{pmatrix}, \gamma \in F_{q^2}, \gamma^q + \gamma = \beta^{q+1}\right\}$$

The group operation is defined as $[1, \beta_1, \gamma_1] \cdot [1, \beta_2, \gamma_2] = [1, \beta_1 + \beta_2, \gamma_1 + \beta_2^q \beta_1 + \gamma_2]$

And $[\beta_1, \gamma_1] \cdot [\beta_2, \gamma_2] = [\beta_1 + \beta_2, \gamma_1 + \beta_2^q \beta_1 + \gamma_2]$.

The factor group $H(P_\infty)/H_1(P_\infty)$ is cyclic of order $q^2 - 1$. Moreover, it was generated by the $\varsigma \in H(P_\infty)$ with $\varsigma(x) = \alpha x$, $\varsigma(y) = \alpha^{q+1} y$.

Another automorphism $\zeta \in H$ is given by $\zeta(x) = x/y$, $\zeta(y) = 1/y$.

The automorphism group $H(P_\infty)$ of the Hermitian function field $Herm|F_{q^2}$ acting on it as $\psi(x), \psi(y)$ has a $ordH(P_\infty) = q^3(q^2-1)$ greater than the order Suzuki group.

A well-known encryption scheme based on the automorphism group of the Hermitian function field is presented in [16].

## BASIC ENCRYPTION SCHEME

Let $H(P_\infty)$ be a large group of the $Herm|F_{q^2}$ with the odd characteristic

$$H(P_\infty) = \left\{ \left[\alpha, \beta, \frac{\beta^{q+1}}{2} + \gamma\right] \middle| \alpha \in F_{q^2}^*, \beta \in F_{q^2}, \gamma^q + \gamma = 0 \right\}.$$

The following steps to apply the encryption scheme.

Step 1. Choose the tame logarithmic signatures $\beta_{(1)} = [B_{1(1)},...,B_{s(1)}] = (b_{ij})_{(1)} = S(1, b_{ij(1)}, b_{ij(1)}^{q+1}/2)$ and $\beta_{(2)} = [B_{1(2)},...,B_{s(2)}] = (b_{ij})_{(2)} = S(1, 0, b_{ij(2)})$ of type $(r_{1(k)},...,r_{s(k)})$, $i = \overline{1,s(k)}$, $j = \overline{1,r_{i(k)}}$, $k=1,2$ $b_{ij(1)} \in F_{q^2}$, $b_{ij(2)} \in F_q \subset F_{q^2}$.

Step 2. Select the random covers $a_{(1)} = [A_{1(1)},...,A_{s(1)}] = (\alpha_{ij})_{(1)} = S\left(a_{ij(1)_1}, a_{ij(1)_2}, (a_{ij(1)_2})^{q+1}/2\right)$ and

$a_{(2)} = [A_{1(2)},...,A_{s(2)}] = (a_{ij})_{(2)} = S\left(a_{ij(2)_1}, a_{ij(2)_2}, (a_{ij(2)_2})^{q+1}/2 + a_{ij(2)_3}\right)$

of the same type as $b_{(1)}$, $b_{(2)}$ where $a_{ij} \in H(P_\infty)$.

Step 3. Choose $\tau_{0(l)}, \tau_{1(l)},...,\tau_{s(l)} \in H(P_\infty) \setminus Z$, $\tau_{i(l)} = S\left(\tau_{i(l)_1}, \tau_{i(l)_2}, (\tau_{i(l)_2})^{q+1}/2\right)$, $t_{i(l)_k} \in F^\times$, $i = \overline{0,s(l)}$, $l=1,2$. Let's $\tau_{s(1)} = \tau_{0(2)}$.

Step 4. Construct a homomorphism $f_1$ defined by $f_1\left(S(a_1, a_2, a_2^{q+1}/2)\right) = S(1, a_2, a_2^{q+1}/2)$

Step 5. Calculating $g_{(1)} = [g_{1(1)},...,g_{s(1)}] = \tau_{(i-1)(1)}^{-1} f_1\left((w_{ij})_{(1)}\right)(v_{ij})_{(1)} \tau_{i(1)}$, $i = \overline{1,s(1)}$, $j = \overline{1,r_{i(1)}}$.

Step 6. Construct a homomorphism $f_2\left(S(a_1, a_2, a_2^{q+1}/2)\right) = S(1, 0, a_2)$.

Step 7. Calculating $g_{(2)} = [g_{1(2)},...,g_{s(2)}] = \tau_{(i-1)(2)}^{-1} f_2\left((a_{ij})_{(2)}\right)(b_{ij})_{(2)} \tau_{i(2)}$, $i = \overline{1,s(2)}$, $j = \overline{1,r_{i(2)}}$.

We got the public key $[f_1, f_2, (a_l, g_l)]$, and a private key $[b_{(l)}, (\tau_{0(l)},...,\tau_{s(l)})]$, $l = \overline{1,2}$.

Let's encrypt. Let $x = S(x_1, x_2, x_3)$ be the message, $x \in H(P_\infty)$.

Step 8. Choose a random $Q = (Q_1, Q_2)$, $Q_1 \in Z_{|F_{q^2}|}$, $Q_2 \in Z_{|\mathbb{Z}|}$.

Step 9. Calculating $y_1 = a'(Q) \cdot x = a_1'(Q_1) \cdot a_2'(Q_2) \cdot x$,

$y_2 = g'(Q) = g_1'(Q_1) \cdot g_2'(Q_2) = S(*, a_{(1)_1}(Q_1) + b_{(1)}(Q_1) + *, a_{(2)_1}(Q_2) + b_{(2)}(Q_2) + *)$.

Cross-calculations of $\tau_{0(l)}, ..., \tau_{s(l)}$ are used to defined $(*)$ components. And it's used for a third coordinate to be added to the product of $a_{(1)_1}(Q_1) + b_{(1)}(Q_1)$.

Step 10. Calculating $y_3 = f_1(a_1'(Q_1)) = S(1, a_{(1)_2}(Q_1), *)$, $y_4 = f_2(a_2'(Q_2)) = S(1, 0, a_{(2)_2}(Q_2))$.

We got vectors $(y_1, y_2, y_3, y_4)$ of the message $x$.

To decrypt a message $x$, we need to restore random numbers $Q = (Q_1, Q_2)$.

Step 11. Calculating $D^{(1)}(Q_1, Q_2) = \tau_{0(1)} y_2 \tau_{s(2)}^{-1} = S(1, a_{(1)_1}(Q_1) + b_{(1)}(Q_1), a_{(2)_1}(Q_2) + b_{(2)}(Q_2) + *)$,

$D^*(Q) = y_3^{-1} D^{(1)}(Q_1, Q_2) = S(1, b_{(1)}(Q_1), a_{(2)_2}(Q_2) + b_{(2)}(Q_2))$.

Restore $Q_1$ with $b_{(1)}(Q_1)$ using $b_{(1)}(Q_1)^{-1}$, because $\beta$ is simple.

For further calculations, it is necessary to remove the components of the arrays $\gamma_1'(Q_1)$ from the ciphertext $y_2$.

Step 12. Calculating $y_2^{(1)} = \gamma_1'(Q_1)^{-1} y_2 = S(*, *, a_{(2)_2}(Q_2) + b_{(2)}(Q_2) + *)$,

$D^{(2)}(Q_2) = \tau_{0(2)} y_2^{(1)} \tau_{s(2)}^{-1} = S(1, 0, a_{(2)_2}(Q_2) + b_{(2)}(Q_2))$, $D^*(Q) = D^{(2)}(Q_2) y_4^{-1} = S(1, 0, b_{(2)}(Q_2))$.

Restore $Q_2$ with $b_{(2)}(Q_2)$ using $b_{(2)}(Q_2)^{-1}$.

We obtain the recovery of $Q = (Q_1, Q_2)$ and the message $x$ from $y_1$.

The correctness of such an implementation is shown in [16]. The considered encryption has several significant disadvantages.

In the encryption algorithm, the keys $Q = (Q_1, Q_2)$ are loosely coupled and allow for a sequential key recovery attack. Key $Q_1$ recovery is possible with brute force attack. Brute force can be performed based on computation $\alpha_1'(Q_1')$ followed by the comparison $y_3$ of the value in the coordinate $a_{(1)_2}(Q_1)$ since $y_3 = f_1(a_1'(Q_1)) = S(1, a_{(1)_2}(Q_1), *)$ searching and finding $Q_1'$ do not depend on the value $Q_2$. Key recovery $Q_2$ is possible through calculation $\alpha_2'(Q_2')$ and comparison within the coordinate $a_{(2)_2}(Q_2)$ -> $y_4 = f_2(a_2'(Q_2)) = S(1, 0, a_{(2)_2}(Q_2))$.

In this case, the complexity of the attack on the keys is equal to $q^2 + q$.

## PROPOSED ENCRYPTION SCHEME IMPROVEMENT

We have improved the encryption algorithm within homomorphic encryption. In this case, the complexity of the key recovery attack will be determined by an exhaustive search over the entire group.

Let's take a look at the basic steps of encryption.

We fix a large group $H(P_\infty)$ of the $Herm|F_{q^2}$ with the odd characteristic.

The group operation is defined as $S(a_1,b_1,c_1) \cdot S(a_2,b_2,c_2) = S(a_1 a_2, a_2 b_1 + b_2, a_2^{q+1} c_1 + a_2 b_2^q b_1 + c_2)$

The identity is $[1,0,0]$ and the inverse of $[\alpha, \beta, \gamma]$ is equal to

$$S(a,b,c)^{-1} = S(a^{-1}, -a^{-1}b, a^{-(q+1)}c^q).$$

$H(P_\infty)$ can be represented more simply by due to the off characteristic:

$$H(P_\infty)\left\{\left[a, b, \frac{b^{q+1}}{2} + c\right] \middle| a \in F_{q^2}^*, b \in F_{q^2} \text{ and } c^q + c = 0\right\}$$

Just show if the generating element of the field is $\gamma$, then the equation $c^q + c = 0$ has solutions $c_i = \gamma^{(q+1)/2 + i(q+1)}$, $i = \overline{0, q-1}$.

## KEY GENERATION STAGE

Step 1. We construct manual logarithmic signatures $\beta_{(k)}$ and random covers

$$a_{(k)} = [A_{1(k)}, ..., A_{s(k)}] = S\left(a_{ij(k)_1}, a_{ij(k)_2}, (a_{ij(k)_2})^{q+1}/2 + a_{ij(k)_3}\right)$$

and $w_{(k)} = [W_{1(k)}, ..., W_{s(k)}] = S\left(1, w_{ij(k)_2}, (w_{ij(k)_2})^{q+1}/2 + w_{ij(k)_3}\right)$ of the same type as $b_{(k)}$, where $a_{ij} \in H(P_\infty)$, $w_{ij} \in H(P_\infty)$, $i = \overline{0, s(k)}$, $j = \overline{1, r_{i(k)}}$, $k = 1, 2$.

Step 2. Generate random $t_{0(k)}, ..., t_{s(k)} \in H(P_\infty) \setminus Z$, $t_{i(k)} = S\left(t_{i(k)_1}, t_{i(l)_2}, (t_{i(k)_2})^{q+1}/2 + t_{i(k)_3}\right)$, $t_{ij(k)} \in F_{q^2}^\times$, $i = \overline{0, s(k)}$, $k = 1, 2$.

Step 3. Choose $\tau_{0(l)}, \tau_{1(l)}, ..., \tau_{s(l)} \in H(P_\infty) \setminus Z$, $\tau_{i(k)} = S\left(\tau_{i(k)_1}, \tau_{i(k)_2}, (\tau_{i(k)_2})^{q+1}/2 + \tau_{i(k)_3}\right)$, $\tau_{i(l)_k} \in F_{q^2}^\times$, $i = \overline{0, s(k)}$, $k = 1, 2$. Let's $t_{s(k-1)} = t_{0(k)}$, $\tau_{s(k-1)} = \tau_{0(k)}$, $k = 1, 2$.

Step 4. Define an additional group operation

$$S(a_1, b_1, c_1) \circ S(a_2, b_2, c_2) = S(a_1 a_2, a_2 b_1 + b_2, a_2^{q+1} c_1 + c_2)$$

The inverse element is $\bar{S}(a,b,c)^{-1} = S(a^{-1}, -a^{-1}b, -a^{-(q_0+1)}c)$.

Let $f(e)$ be a homomorphic cryptographic transformation concerning addition $f(a+b) = f(a) + f(b)$ $e, a, b \in F_q$ and the corresponding inverse transformation $\hat{f}(e) = e$.

Step 5. We calculate the covering of the logarithmic signatures

$$h_{(1)} = [h_{1(1)}, ..., h_{s(1)}] = t_{(i-1)(1)}^{-1} \cdot (w_{ij})_{(1)} \cdot (b_{ij})_{(1)} \cdot t_{i(1)}, \quad h_{(2)} = [h_{1(2)}, ..., h_{s(2)}] = \bar{t}_{(i-1)(2)}^{-1} \circ (w_{ij})_{(2)} \circ (b_{ij})_{(2)} \circ t_{i(2)}$$

and coverings of the homomorphic cryptographic transformation.

$$g_{(1)} = [g_{1(1)}, ..., g_{s(1)}] = \tau_{(i-1)(1)}^{-1} \cdot f(w_{ij})_{(1)} \cdot \tau_{i(1)}, \quad g_{(2)} = [g_{1(2)}, ..., g_{s(2)}] = \bar{\tau}_{(i-1)(2)}^{-1} \circ f(w_{ij})_{(2)} \circ \tau_{i(2)},$$

where $f(w_{ij}) = S\left(1, f(w_{ij(k)_2}), f\left((w_{ij(k)_2})^{q+1}/2\right) + f(w_{ij(k)_3})\right)$, $i = \overline{1, s(k)}$, $j = \overline{1, r_{i(k)}}$, $k = 1, 2$.

Let's $\hat{f}(w_{ij})$ is an inverse transformation in regards to $f(w_{ij})$.

We get the public key $(a_k, h_k, g_k)$, and a private key $[f, \beta_{(k)}, t_{i(k)}, \tau_{i(k)}]$, $i = \overline{0, s(k)}$, $k = 1, 2$.

## ENCRYPTION STAGE

We have message $x \in H(P_\infty)$, $x = S(x_1, x_2, x_3)$, and the public key $(a_k, h_k, g_k)$, $k = 1, 2$ as input data.

Step 6. Choose a random $Q = (Q_1, Q_2)$, $Q_1 \in Z_{|F_{q^2}|}$, $Q_2 \in Z_{|\mathbb{Z}|}$.

Step 7. Calculating the ciphertext $y_1, y_2, y_3$

$$y_1 = a'(Q) \cdot x = a_1'(Q_1) \cdot a_2'(Q_2) \cdot x,$$

$$y_2 = h'(Q) = h_1'(Q_1) \cdot h_2'(Q_2) = S\left(*, \sum_{i=1, j=Q_{i(1)}}^{s(1)} (w_{ij(1)_2} + \beta_{ij(1)}) + *, \sum_{i=1, j=Q_{i(2)}}^{s(2)} (w_{ij(2)_3} + \beta_{ij(2)}) + *\right),$$

Here, the $(*)$ components are determined by cross-calculations in the group operation of the product of $t_{0(k)}, ..., t_{s(k)}$ and the product of $w_{(k)}(Q_k), \beta_{(k)}(Q_k)$.

$$y_3 = g'(Q) = g_1'(Q_1) \cdot g_2'(Q_2) = S\left(*, \sum_{k=1}^{2} \sum_{i=1, j=Q_{i(k)}}^{s(k)} w_{ij(k)_2} + *, \sum_{k=1}^{2} \sum_{i=1, j=Q_{i(k)}}^{s(k)} w_{ij(k)_3} + *\right)$$

Here, the $(*)$ components are determined by cross-calculations in the group operation of the product of $\tau_{0(k)}, ..., \tau_{s(k)}$ and the product of $w_{(k)}(Q_k)$.

As a result we get a ciphertext $(y_1, y_2, y_3)$ of the message $x$.

## DECRYPTION STAGE

We have a ciphertext $(y_1, y_2, y_3)$ and private key $[f, \beta_{(k)}, t_{i(k)}, \tau_{i(k)}]$, $i = \overline{0, s(k)}$, $k = 1, 2$.

To decrypt a message $x$, we need to restore random numbers $Q = (Q_1, Q_2)$.

Step 8. Calculating $D(Q) = t_{0(1)} y_2 \circ \bar{t}_{s(3)}^{-1}$, $G(Q) = \tau_{0(1)} y_3 \circ \bar{\tau}_{s(3)}^{-1}$,

$$D(Q)' = D(Q) \cdot \hat{f}(G(Q))^{-1} = S(1, \sum_{i=1, j=R_{i(1)}}^{s(1)} \beta_{ij(1)}, *)$$

Step 9. Restore $Q_1$ with $\beta_{(1)}(Q_1) = \sum_{i=1, j=Q_{i(1)}}^{s(1)} \beta_{ij(1)}$ using $\beta_{(1)}(Q_1)^{-1}$, because $\beta_{(1)}$ is simple.

For further calculation, it is necessary to remove the component $h_1'(Q_1)$ from $y_2$ and $g_1'(Q_1)$ from $y_3$.

Step 10. Calculating $y_2^{(1)} = h_1'(Q_1)^{-1} \cdot y_2$, $y_3^{(1)} = g_1'(Q_1)^{-1} \cdot y_3$, $D(Q)^{(1)} = t_{0(2)} \circ y_2^{(1)} \circ \bar{t}_{s(3)}^{-1}$,

$G(Q)^{(1)} = \tau_{0(2)} \circ y_3^{(1)} \circ \bar{\tau}_{s(3)}^{-1}$, $D(Q)'' = D(Q)^{(1)} \circ \hat{f}(G(Q)^{(1)})^{-1} = S(1, 0, \sum_{i=1, j=R_{i(2)}}^{s(2)} \beta_{ij(2)})$

and restore $Q_2$ with $\beta_{(2)}(Q_2) = \sum_{i=1, j=R_{i(2)}}^{s(2)} \beta_{ij(2)}$ using $\beta_{(2)}(Q_2)^{-1}$, because $\beta_{(2)}$ is simple.

Recovery the message $x = a'(Q_1', Q_2')^{-1} \cdot y_1$.

## SECURITY ANALYSIS

Let's look at the main aspects of brute-force key recovery and algorithm attacks.

Attack 1. Brute force attack on the ciphertext $y_1$.

By choosing $Q = (Q_1, Q_2)$, the attacker will try to decipher the text $y_1 = a'(Q) \cdot x$. The covers $a_{(k)}$ are selected at random and $a'(Q)$ the value is determined by multiplication within a group. The resulting $a'(Q)$ vector depends on all $a_k'(Q_k)$ components. An enumeration of all values of the $Q = (Q_1, Q_2)$ key has an estimate of the complexity of $q^3$. For a practical attack, the $x$ the message is unknown and has an uncertainty of choice of $q^3$. This makes it impossible for a brute force attack on the key through comparison by the value of $y_1$. If we take an attack model with a known text, then the complexity of the attack will remain the same and equal to $q^3$.

Attack 2. Brute force attack on the ciphertext $y_2$.

The attacker selects such values of $Q_1'$ and $Q_2'$ for which the calculated $y_2(Q_1',Q_2')$ coincides with the true value of the $y_2(Q_1,Q_2)$. The keys $Q=(Q_1,Q_2)$ are related, and changing any of them results in a change $y_2$. A brute force attack on a key has a difficulty equal to $q^l$.

Attack 3. Brute force attack on the ciphertext $y_3$.

The attacker selects the values of the $Q=(Q_1,Q_2)$ keys, calculate the $y_3(Q_1',Q_2')$ and compare them with the desired $y_3(Q_1,Q_2)$. The $y_3$ value for all coordinates depends on the values $w_{ij(k)_2}$, $w_{ij(k)_3}$ of the vectors of the $W_{1(k)},...,W_{s(k)}$ arrays selected by the key. So the keys $Q_1, Q_2$ are linked and changing any of them leads to a change in the $y_3$. Thus, a brute force attack on a key has a difficulty equal to $q^l$.

Attack 4. Brute force attack on the vectors $(t_{0(k)},...,t_{s(k)})$ and $(\tau_{0(k)}, \tau_{1(k)},...,\tau_{s(k)})$.

The brute force attack on $(t_{0(k)},...,t_{s(k)})$ is a general for the MST cryptosystems and for the calculation in the field $F_q$ over the group center $Z(G)$ has an optimistic complexity estimation equal to $q$. For the proposed algorithm all calculations are executed on the whole group $|A_l(n,\theta)|=q^5$ and in such cases the complexity of the brute force attack on $(t_{0(k)},...,t_{s(k)})$ and $(\tau_{0(k)}, \tau_{1(k)},...,\tau_{s(k)})$ will be equal to $q^5$.

Attack 5 on the algorithm.

The attack on the implementation algorithm of the encryption scheme based on the automorphism group of the Hermitian function field has many different aspects. Practical attacks look at the design features of logarithmic signatures and random coverings. The attacker builds attacks based on known texts and random coverings. Known solutions for constructing secret cryptosystems based on the use of aperiodic logarithmic signatures. In the proposed cryptosystem with homomorphic encryption, random coverings are a secret for the cryptanalyst. In this case, the known attacks based on the weakness of logarithmic signatures are impossible.

## CONCLUSION

MST cryptosystems leveraging the automorphism group of the Hermitian function field demonstrate significant advantages in both security and implementation efficiency compared to alternative approaches. This framework enables the construction of high-security cryptosystems while performing group computations over small finite fields. The integration of homomorphic encryption with random coverings in logarithmic signatures provides robust protection against established attack vectors targeting logarithmic signature implementations. Furthermore, this approach allows for the utilization of secure logarithmic signatures with straightforward design, resulting in reduced computational overhead for the cryptosystem's general parameters. Consequently, the proposed homomorphically enhanced cryptosystem represents a promising candidate for post-quantum cryptographic applications.